\documentclass[doublespacing]{elsart}

\usepackage{graphicx}
\usepackage{amssymb}

\begin{document}

\begin{frontmatter}



\title{Effects of neighbourhood size and connectivity on the spatial
Continuous Prisoner's Dilemma}


\author[labelp,labelt]{Margarita Ifti}
\author[labele]{Timothy Killingback}
\author[labelz]{Michael Doebeli}

\address[labelp]{Department of Physics and Astronomy, University of 
British Columbia, Vancouver, BC, Canada V6T 1Z1}
\address[labele]{Ecology and Evolution, ETH Z\"urich, 8092 Z\"urich, 
Switzerland}
\address[labelz]{Department of Mathematics and Zoology, University of 
British Columbia, Vancouver, BC, Canada V6T 1Z4}
\address[labelt]{Permanent address: Department of Physics, University
of Tirana, Tirana, Albania}

\begin{abstract}

The Prisoner's Dilemma, a 2-person game in which the players can either
cooperate or defect, is a common paradigm for studying the evolution of 
cooperation. In real situations cooperation is almost never all or 
nothing. This observation is the motivation for the Continuous Prisoner's 
Dilemma, in which individuals exhibit variable degrees of cooperation. 
It is known that in the presence of spatial structure, when individuals 
``play against'' (i.e. interact with) their neighbours, and ``compare to'' 
(``learn from'') them, cooperative investments can evolve to considerable 
levels. Here we examine the effect of increasing the neighbourhood size: 
we find that the mean-field limit of no cooperation is reached for a 
critical neighbourhood size of about five neighbours on each side in a 
Moore neighbourhood, which does not depend on the size of the spatial 
lattice. We also find the related result that in a network of players, the 
critical average degree (number of neighbours) of nodes for which 
defection is the final state does not depend on network size, but only on 
the network topology. This critical average degree is considerably (about 
ten times) higher for clustered (social) networks, than for distributed 
random networks. This result strengthens the argument that clustering is 
the mechanism which makes the development and maintenance of the 
cooperation possible. In the lattice topology, it is observed that when 
the neighbourhood sizes for ``interacting'' and ``learning'' differ by 
more than 0.5, cooperation is not sustainable, even for neighbourhood 
sizes that are below the mean-field limit of defection. We also study the 
evolution of neighbourhood sizes, as well as investment level. Here we 
observe that the series of the interaction and learning neighbourhoods 
converge, and a final cooperative state with considerable levels of 
average investment is achieved.

\end{abstract}

\begin{keyword}
Prisoner's Dilemma \sep continuous \sep spatial \sep lattice \sep network

% \PACS 
\end{keyword}
\end{frontmatter}

% main text

\section{Introduction}

The origin of cooperation is a fundamental problem in evolutionary biology.
Cooperation is essential in the functioning of almost every known biological 
system~(Hamilton (1964a), Hamilton (1964b), Trivers (1971), Dugatkin 
(1997)). For example, according to Eigen \& Schuster (1979), Michod (1983),
and Maynard Smith \& Szathm\'ary (1995), early replicating molecules may 
have cooperated to form larger entities which could encode more 
information. Also, the transition from free-living single-cell protists to 
multicellular organisms seems to have depended on cooperation~(Maynard 
Smith \& Szathm\'ary (1995), Buss (1987)).

It is however, difficult to explain why individuals should cooperate. In 
the traditional Prisoner's Dilemma model of cooperation, defecting 
individuals always have a higher fitness than cooperators. Cooperation is 
not an evolutionary stable strategy, because it can be invaded by 
defectors. Hence, the emergence of cooperation is generally assumed to 
require repeated play (with memory) and strategies such as Tit for Tat, or 
``tags''~(Axelrod (1984), Guttman (1996), Lindgren \& Nordahl (1994), 
Miller (1996)). The work of Nowak \& May (1992) showed that placing 
ensembles of cooperators and defectors on a lattice generates changing 
spatial patterns, in which both cooperators and defectors persist 
indefinitely. The introduction of spatial structure changes the picture 
from the mean-field result in which defection always wins to a final state 
with both cooperators and defectors present. Similar results were obtained 
by Epstein (1998), who introduced the Demographic Prisoner's Dilemma, 
in which the individuals have a fixed strategy (which is their phenotype), 
but are placed in a spatially structured lattice world. Epstein (1998) 
found that regions of cooperation persisted in this spatial model. The 
studies of Nakamaru et al. (1997), Iwasa et al. (1998), Nakamaru et al. 
(1998), and Irwin and Taylor (2001) showed that spatially structured 
models, such as the lattice model, produce the clumping of the cooperative 
players, and then enables them to invade a population of defectors, but the 
spatial structure also encourages the evolution of spiteful behaviour. 
These models consider the invasiveness and stability of fully developed, 
highly cooperative interactions.

The gradual evolution of cooperation from an initially selfish state 
represents a more plausible evolutionary scenario. It is then more natural 
to consider models in which several degrees of cooperation are 
possible~(Doebeli \& Knowlton (1998), Roberts \& Sherratt (1998), Wahl 
\& Nowak (1999a), Wahl \& Nowak (1999b), Szab\'o \& Hauert (2002a), 
Szab\'o \& Hauert (2002b), Killingback \& Doebeli (2002)). When we take 
into account the possibility of variable levels of cooperation, we can 
study the crucial issue of how cooperation can gradually evolve from a 
non-cooperative initial state. Roberts \& Sherratt (1998) considered a 
``raise-the-stakes'' strategy for the iterated PD, and showed that it 
invades and is stable against a number of alternative strategies. Doebeli 
\& Knowlton (1998) considered interspecific symbiosis in the context of 
iterated asymmetric PD, and concluded that such interactions could 
increase in extent and frequency if the populations are spatially 
structured. In this model, strategies with very low levels of cooperation 
can gradually evolve to much more cooperative strategies. The end result 
is a high degree of mutualism between pairs of interacting individuals 
that belong to different species.

Killingback et al. (1999) extended the classical Prisoner's Dilemma, 
introducing a model of cooperation which is based on the concept of 
investment, and develops further the ideas of Doebeli \& Knowlton (1998). 
This evolutionary game is called Continuous Prisoner's Dilemma (CPD). 
Killingback et al. (1999) showed that intraspecific cooperation easily 
evolves from very low levels, and is sustained, with fluctuations, at 
relatively high levels, when the game is played in spatially structured 
populations. Killingback et al. (1999) assume that individuals play 
against their immediate neighbours, and also compare their payoffs to 
those of the same individual neighbours. It is important to know how 
robust are the results obtained by Killingback et al. (1999) when these 
assumptions are relaxed, i.e. when individuals are allowed to play against 
more distant neighbours (than their nearest ones), and then compare their 
payoffs to those of a different group of neighbours, which may be larger 
or smaller than the ones included in the first interaction neighbourhood. 
Also, Killingback et al. (1999) conjecture that clustering is the 
mechanism that allows the establishment and maintenance of a cooperative 
state. To investigate the validity of this hypothesis, we studied the 
behaviour of the CPD game on different topologies, such as networks with 
different clustering properties.

\section{The Basic Model: The Continuous Prisoner's Dilemma}

\begin{figure}
\includegraphics{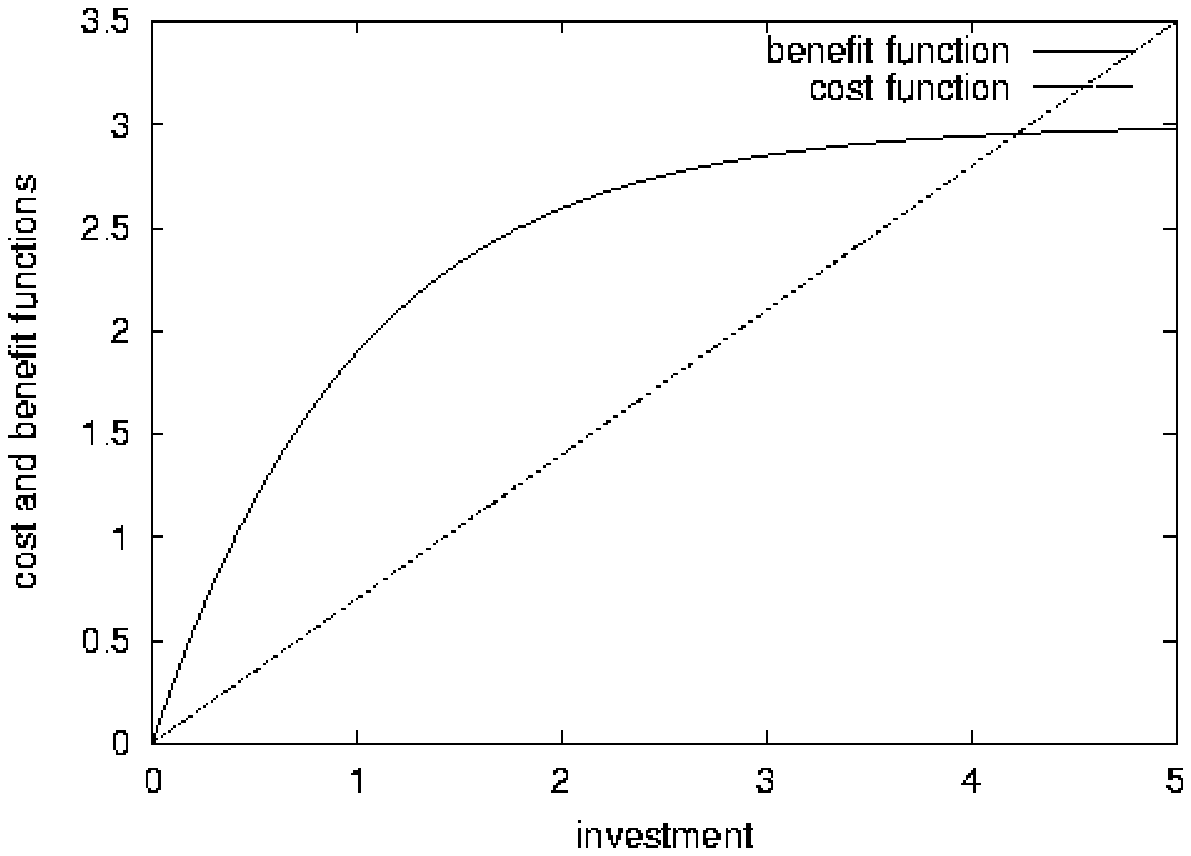}
\caption{Possible benefit and cost functions for the CPD game.}
\label{fig:fig1}
\end{figure}

The Continuous Prisoner's Dilemma (CPD) game between two individuals is
based on the assumption that each of them makes an investment (which can 
take any non-negative real value). Making an investment $I$ has the effect 
of reducing the fitness of the individual who makes it by ``the cost'' 
$C(I)$ and increasing the fitness of the beneficiary by ``the benefit'' 
$B(I)$. So, if two individuals 1 and 2, play against each other and make 
investments $I_1$ and $I_2$, the payoff of 1 is $B(I_2)-C(I_1)$ and that 
of 2 is $B(I_1)-C(I_2)$. Possible benefit and cost functions are shown in 
Fig.~\ref{fig:fig1}. Cost and benefit functions of this type are typical 
of what might be expected in a real biological situation, such as those 
discussed by Hart \& Hart (1992) and Wilkinson (1984). The common 
feature of the functions $C(I)$ and $B(I)$ is that $B(I)>C(I)$ for a range 
of the argument $0<I<I_{max}$ (for $I>I_{max}$ the cost is higher than the 
benefit, so it does not make sense for a player to invest an amount higher 
than $I_{max}$). Limiting the investment levels to just a pair of 
investment amounts would bring us back to the standard Prisoner's Dilemma.

In the absence of an additional structure, i.e. in the mean-field
approximation, or the well-mixed system, the zero-investment strategy is 
again the winning one. Starting at any level, investments will gradually 
evolve to zero, since the defectors will benefit from the investment of 
the cooperators without bearing the costs. Using adaptive dynamics, one 
can see that the zero investment strategy is the evolutionary stable
strategy.

\section{CPD in a Lattice}

Killingback et al. (1999) introduced spatial structure into the model, 
following the general approach of spatial evolutionary game theory 
(Axelrod (1984), Nowak \& May (1992), Killingback \& Doebeli (1996)). The 
individuals are placed in the cells of a 2D square lattice. Each 
individual makes an investment, and interacts with the individuals within 
a Moore neighbourhood of hers. A Moore neighbourhood of, e.g. size one, 
includes a cell's eight immediate neighbours: north, northeast, east, 
southeast, south, southwest, west, northwest, i.e. a ``ring'' of 
``thickness'' one neighbour in all directions; that of size 2 will include 
all individuals within a ``ring'' of ``thickness'' two neighbours in all 
directions, etc. In this paper we will refer to the thickness of the 
interaction neighbourhood as the {\bf neighbourhood parameter}. The 
individual will get payoffs as prescribed by the rules of the game, when 
playing against each of the neighbours within her interaction 
neighbourhood. Her fitness is then the sum of the payoffs she gets from 
playing against all her neighbours. (In the process, the fitness of the 
neighbours is calculated too, as the sum of the payoffs they get from 
playing against the individuals within their own interaction 
neighbourhoods.) She then compares her payoff to those of the individuals 
within her learning neighbourhood, whose size (thickness) may be the same 
or different from that of the interaction neighbourhood. In traditional 
spatial evolutionary game theory, this distinction between the 
individual's interaction and learning neighbourhood is not made, silently 
assuming that they are the same. In this paper, we will refer to the 
thickness of the learning neighbourhood as the {\bf dispersal parameter}. 
Our focal individual then identifies the neighbour with the highest 
fitness, and adopts that neighbour's strategy, by changing the amount of 
her investment to that of the neighbour's. At this stage the investment 
level of the player can be mutated at a fixed probability. This 
corresponds to an economic scenario in which the agents learn (with 
occasional errors) from their more successful partners, or to an 
evolutionary scenario in which the more successful phenotypes replace the 
less successful ones (with mutations). Every few generations the average 
investment per individual is calculated. 

Killingback et al. (1999) simulated the CPD game in a square lattice with 
neighbourhood sizes for both interacting and learning (i.e. both the 
neighbourhood and dispersal parameters) set to one. They observed that the 
average investment increases from an almost zero starting value, to a 
significant final level of investment. Its average value for a cost 
function $C(I)=0.7 \cdot I$ and benefit function $B(I)=8 \cdot (1-e^{-I})$, 
in a $70 \times 70$ lattice, with periodic boundary conditions, is around 
$1.05$. (The probability of mutations is set to $0.01$, and they follow a 
Gaussian curve with centre at the individual's investment value, and width 
10 percent of its peak.) The average investment amount is maintained close 
to $1.05$ by the dynamics of the spatial system. It is smaller than the 
maximum investment $I_{max}$, but much larger than the initial maximum 
investment amount of 0.0001. The key factor that causes a state with 
considerable average investment levels to be established is the fact that 
higher investing individuals benefit from clustering together~(Killingback 
et al. (1999)). This result agrees with those of Nakamaru et al. (1997), 
Iwasa et al. (1998), Nakamaru et al. (1998), and Irwin and Taylor (2001).

\begin{figure}
\includegraphics{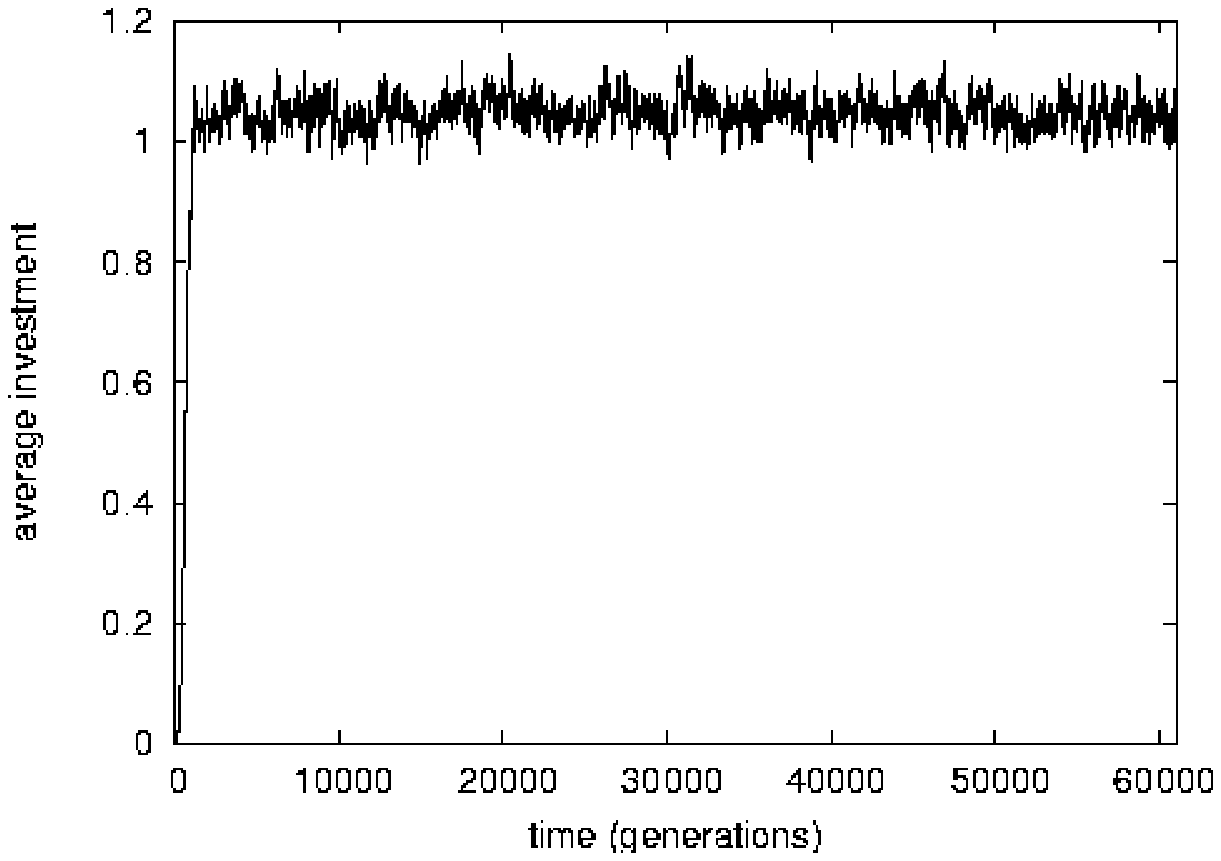}
\includegraphics{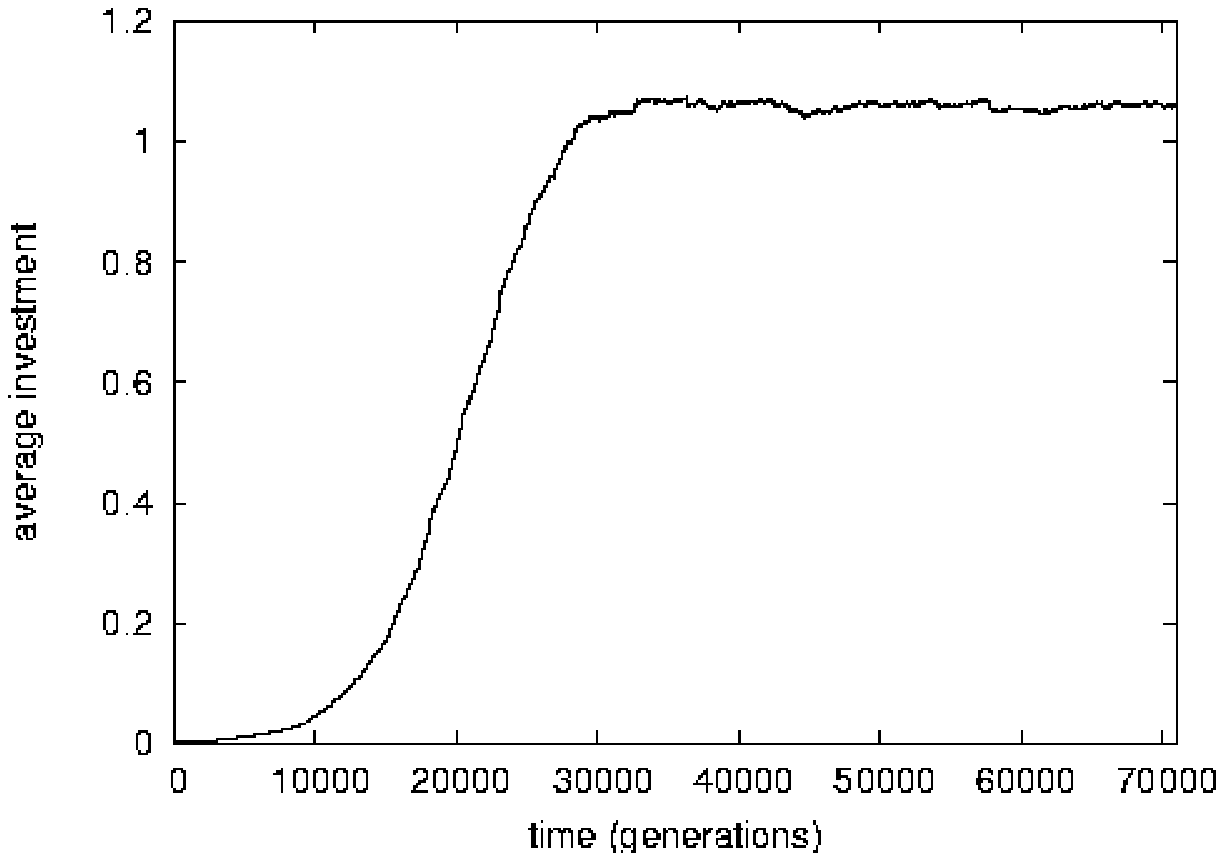}
\caption{The evolution of average investment with time when we only 
consider the nearest neighbours for interaction-control run. The top 
graphh shows a time series obtained for a high mutation rate of 5\%, the 
bottom graph corresponds to a mutation rate of 0.5\%.}
\label{fig:11}
\end{figure}

We introduced ``real-time'' evolution as follows: every individual gets
her own ``clock'' which tells her when to ``wake up'', look around, play
against the neighbours within the interaction neighbourhood, compare her 
payoff to that of the neighbours within the learning neighbourhood, and
revise her strategy. The ``wake up'' times are generated as $-\ln (rn)$
where $rn$ is a random number with uniform distribution in $[0,1]$, which 
guarantees that the events are independent~(Gibson \& Bruck (2000)). (This 
corresponds to asynchronous updating, but with a different distribution of 
``wake up'' events.) We use the same lattice size, cost and benefit 
functions, and mutation rate as Killingback et al. (1999). After a 
sufficiently long time (which is of the order of 2~000--3~000 generations), 
the ensemble again reaches a state with a considerable degree of 
cooperativity, and the average investment level is about the same as that 
obtained by Killingback et al. (1999).  A typical evolutionary outcome is 
shown in Fig.~\ref{fig:11} (top graph). 

Here we can take a moment to discuss the role of mutations. The top graph 
in Fig.~\ref{fig:11} corresponds to a very high mutation rate of 5\%, with 
Gaussian distribution; the width of the Gaussian is set at 10\% of its 
height. In Fig.~\ref{fig:11}, in the bottom graph, we show a time series 
obtained for a mutation rate of 0.5\%, width of the Gaussian is 1\% of its 
height. It is clear that the mutation rate does not determine the mean 
value of the average investment in the steady state reached by the system; 
it rather influences how fast the steady state is reached (occasional 
``kicks'' by mutations help the average investment grow quickly), and the 
magnitude of fluctuations in steady state.

One general question is: what amount of spatial structure do we need 
(i.e. how big can neighbourhood sizes become) before we reach the 
mean-field (all defection) situation? Keeping a Moore type of 
neighbourhood, with the neighbourhood and dispersal parameters equal to 
each-other (for the moment), we consider the cases where an individual 
plays against everybody within a neighbourhood of radius 2 (24 
individuals), 3 (48 individuals), 4 (80 individuals), and so on. The rules 
of the game, the cost function, benefit function, mutation rate, and width 
of the Gaussian mutation curve, are kept fixed and equal to those used by 
Killingback et al. (1999). For each case, we averaged over 500 realisations 
of the system, and report results obtained on a $70 \times 70$ lattice, 
unless otherwise stated. 

\begin{figure}
\includegraphics{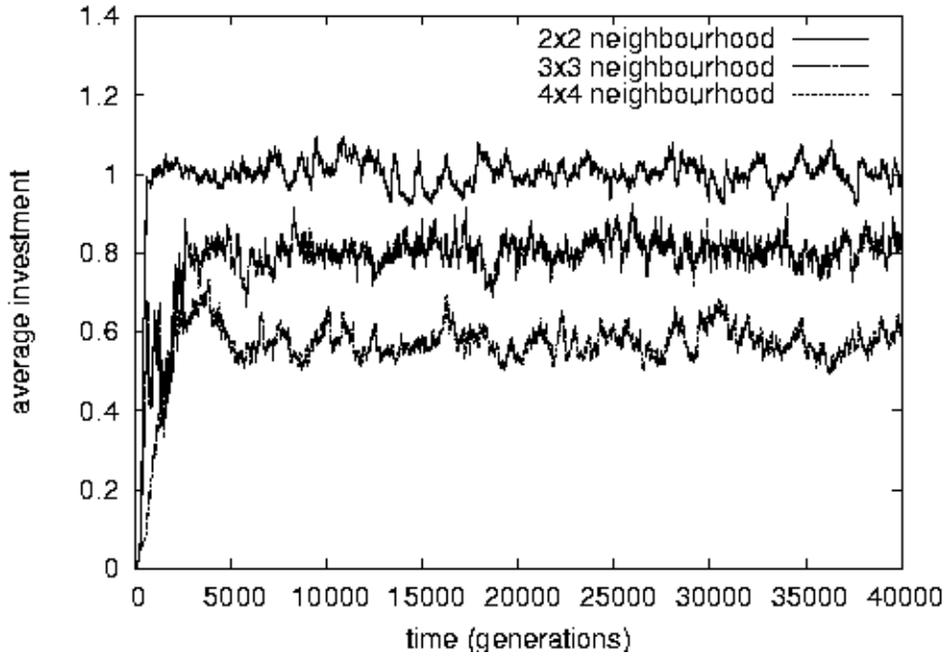}
\caption{Average investment vs. time for the cases when we consider more 
than one neighbour, here the neighbourhood and dispersal parameters are
equal and vary from 2 to 4.} 
\label{fig:nn}
\end{figure}

For neighbourhood and dispersal parameters equal to 2 (i.e. a 
neighbourhood that includes 24 individuals, as opposed to 8 for
neighbourhood parameter 1), the final average investment value is very
close to that obtained when those parameters are 1. When we increase the 
neighbourhood and dispersal parameters to 3, the final average investment 
drops to around 0.85, and it drops even further to about 0.6 for 
parameters' values equal to 4. For larger lattice sizes we get similar 
values of average investment in the steady state. Some typical runs are 
shown in Fig.~\ref{fig:nn} for lattice size $70 \times 70$. 

\begin{figure}
\includegraphics{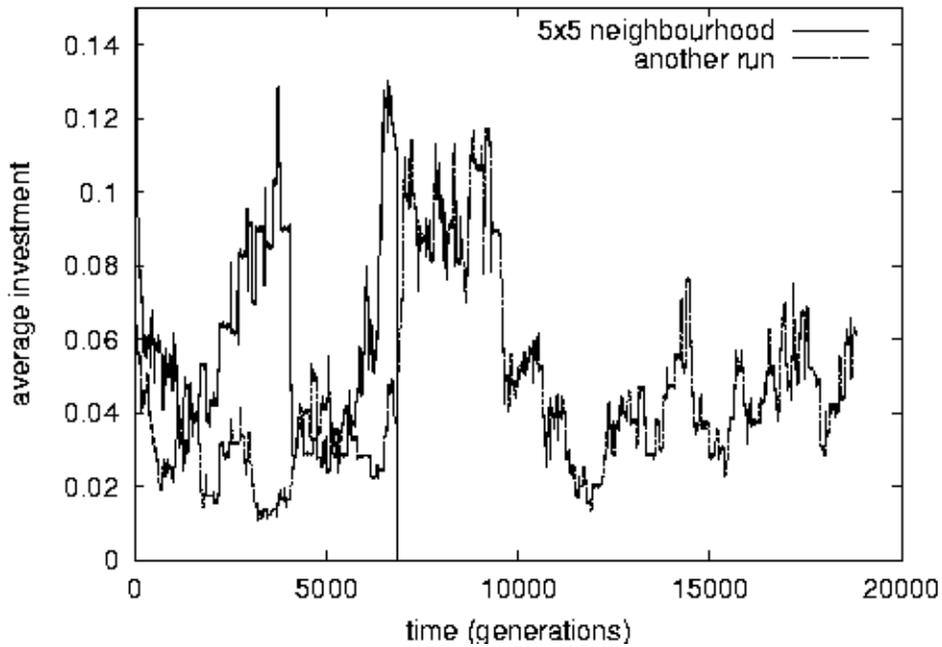}
\caption{In the case when the neighbourhood and dispersal parameter 
become 5, we are at the mean--field limit.}
\label{fig:55}
\end{figure}

The picture changes qualitatively when the neighbourhood and dispersal 
parameters become 5. Even though we start with a relatively large initial 
investment, the final average investment is extremely small (about 0.04,
as opposed to 0.6 for a size four neighbourhood), and can occasionally 
drop to zero under the influence of even moderate fluctuations. This is 
shown in Fig.~\ref{fig:55}. The gradual drop in the value of average 
investment as we increase the neighbourhood size is typical for 
nonequilibrium phase transitions. For larger lattice sizes we get similar 
results for the average investment, but, as expected, the fluctuations are 
lower. We can then state that at this neighbourhood size we have reached 
the mean-field limit, i.e. the picture we would obtain in absence of a 
spatial structure altogether (which for the CPD corresponds to a state 
of pure defection, as discussed above). The value of the neighbourhood 
and dispersal parameters for which the mean-field limit is reached was 
found to be independent of the lattice size, at least for reasonably large 
lattices (the meaning of ``reasonably large'' will become clear in the 
next paragraph). The immediate neighbourhood comprises about 100 
individuals, while the extended one (the neighbours of the neighbours) 
comprises about 300. Furthermore, the results were independent of the 
choice of the boundary conditions; they hold also when periodic boundary 
conditions are not used (and the last row cells have only five neighbours, 
as opposed to eight the other cells have, for a size one neighbourhood, 
and similarly for other neighbourhood sizes).

\begin{figure}
\includegraphics{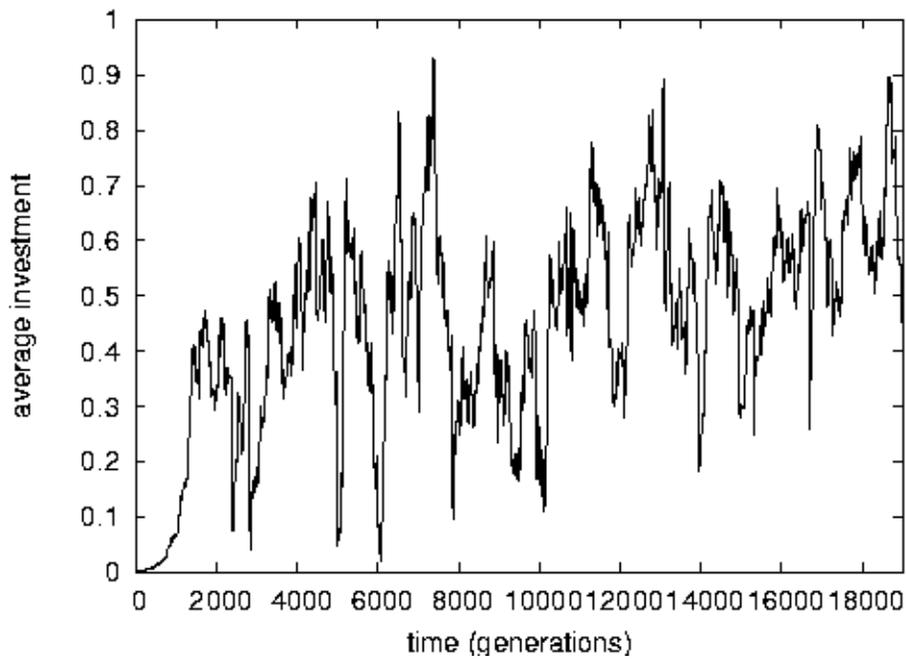}
\caption{The evolution of average investment with time in the case when 
both the neighbourhood and dispersal parameters are 4, but the lattice
size is $30 \times 30$. Note the huge fluctuations, which bring the
investment very close to zero as a result of mutations.}
\label{fig:44sm} 
\end{figure}

Let us now examine the role of the ratio of the neighbourhood size to 
lattice size. In Fig.~\ref{fig:44sm} we show a time series of average 
investment for the case when both the neighbourhood and dispersal 
parameters are 4, and the lattice size is $30 \times 30$. It is important 
to note that for smaller lattices the size of fluctuations can become 
comparable to the average investment amount, even when the latter is 
considerably above zero, and often a mutation is enough to bring the 
average investment amount to zero. The average investment does come 
``dangerously'' close to zero a few times in Fig.~\ref{fig:44sm}, and we 
observed it drop to zero in many other runs not shown here. So, if the 
lattice is too small, the fluctuations are enormous, and one might be 
misled by the result of a particular simulation. (For nonequilibrium phase 
transitions like this one, physicists scale the average investment with the 
lattice size, but this goes beyond the purpose of the present paper.) The 
$70 \times 70$ lattice size we worked with (most of the time) is large 
enough to keep the fluctuations moderate, but also small enough to allow 
our simulations to run within reasonable times. We can conclude, based on 
our simulations, that the neighbourhood size for which we achieve the 
mean-field limit is about 5, and does not depend on the lattice size (at 
least for reasonably large lattices), contrary to the expectations of 
Killingback et al. (1999). The larger neighbourhood sizes ``expose'' the 
players that sit close to the borders of the square to a larger number of 
defector ``outsiders'', whose easily earned large payoffs might ``tempt'' 
our players. The numbers of these ``tempters'' only depend on the 
neighbourhood size. However, the lattice size influences a very important 
factor, the size of the fluctuations, which can make or break the fate of 
the cooperative state.

\section{Variable Neighbourhood Size, the Role of Information, and
Evolution}

The above results were obtained when the size of the interaction 
neighbourhood was the same as that of the learning neighbourhood. It is 
interesting to investigate what happens when this requirement is relaxed, 
and the neighbourhood and dispersal parameters are allowed to be different 
from each-other. For this purpose, we simulated a situation in which our 
agents play against individuals within a certain neighbourhood, but 
compare their payoffs to those of individuals within a neighbourhood that 
is different from their interaction neighbourhood. We observe that, if 
the neighbourhood parameter differs from the dispersal parameter by one or 
more, then cooperation is no longer sustained. Furthermore, the final 
state is defection, independent of whether the neighbourhood parameter is 
larger than the dispersal one, or vice-versa. It appears as if, when the 
neighbourhood parameter is smaller than the dispersal parameter, the 
defectors outside the interaction neighbourhood, but within the learning 
neighbourhood, manage to ``tempt'' the cooperators. While in the case when 
the neighbourhood parameter is larger than the dispersal one, there are 
simply too many defectors in the interaction neighbourhood, taking 
advantage of the cooperators. In the economic context, too little or too 
much information favours defection, and kills the incentive to cooperate.

In the real life, however, there is no rigid separation of neighbourhoods 
from one-another. Individuals interact not only within a strictly defined 
neighbourhood, but rather occasionally include other individuals in their 
sphere of relations. This motivated us to determine what fractional 
difference between the neighbourhood and dispersal parameters is required
for cooperation to be no longer sustained. This can be studied in our 
model by making the neighbourhood and dispersal parameters ``continuous''. 
To do this we constructed a ``fractional size'' neighbourhood in the 
following way: suppose the parameter has the form  $<m.n>$ with $m$ the 
integer part and $<0.n>$ the fractional part. Then for the $(m+1)$-th 
neighbour we generate a random number with uniform distribution in 
$[0,1]$. If the random number is smaller than $<0.n>$, the individual 
plays against this particular neighbour, otherwise the next event in the 
queue is generated. If the neighbour is at the corner, the random number 
is compared to $(0.n)^2$ (i.e. the fractional area of the rectangle). The 
remaining rules of the game are unchanged. 

\begin{figure}
\includegraphics{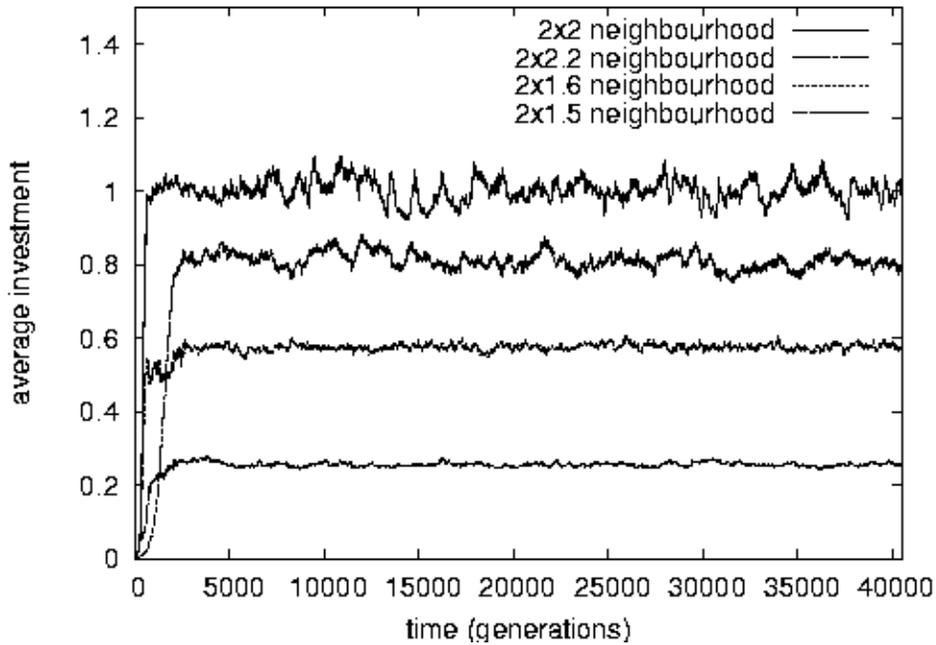}
\caption{When both neighbourhood and dispersal parameters are 
``continuous'', the cooperation dies when their difference is about 0.5. 
All graphs are for neighbourhood parameter equal to 2; the top graph is the 
control one, corresponding to dispersal parameter equal to 2, then follow 
from top to bottom the time series for dispersal parameters equal to 2.2 
(average investment in steady state is the same as for dispersal parameter 
equal to 1.8), then dispersal parameters equal to 1.6 and 1.5. When the 
dispersal parameter is 1.4, the final average investment drops to zero.}
\label{fig:c20} 
\end{figure}

We find in this situation that cooperation only persists for differences 
between parameters up to about 0.5, and breaks down when the difference 
between the neighbourhood and dispersal parameters is 0.5 or larger. We 
have looked at both the case when the neighbourhood parameter is larger or 
smaller than the dispersal parameter. The results are the same: as soon as 
the difference between the two parameters becomes larger than 0.5, 
cooperation vanishes. Some results with different neighbourhood and 
dispersal parameters are shown in Fig.~\ref{fig:c20}. Here we can see that 
the average investment has already dropped to zero, when the difference 
between the neighbourhood and dispersal parameters reaches 0.6, and is 
very small, when that difference is 0.5.

There is no reason why every agent should have the same parameters, so we 
may treat them as ``local'', i.e. every individual has her own 
neighbourhood and dispersal parameters, $n_i$ and $d_i$, which generally 
are not integers. When her turn comes, she plays against individuals within 
a neighbourhood of radius $n_i$ and then compares herself to the 
individuals within the neighbourhood of radius $d_i$. However, there are 
cases when she meets a neighbour $j$, whose dispersal parameter is smaller 
than hers (i.e. he does not ``see'' her, even though she ``sees'' him). In 
such cases she could include him into her learning neighbourhood, comparing 
her payoff to his, and then adopting his investment amount, if he is doing 
better than her. This would correspond to the economics scenario, when the 
focal player can learn from players who do not see her. Otherwise she could 
leave that player who is doing better than her, but cannot see her, out of 
her learning neighbourhood, even though she sees him. This would correspond 
to the evolutionary scenario, where a player can only take over a cell he 
does see. A natural question is: do the two scenarios lead to different 
results?

We consider both possibilities: first when she does compare her payoff to 
that of the neighbour with a smaller dispersal parameter (and, if he is 
doing better, adopts his strategy, i.e. investment amount), and second when 
she does not. For that we had both the neighbourhood and dispersal 
parameters initialized randomly between the values 1 and 4 (which are 
the interesting values, since for larger parameters there is no 
sustainable cooperation). The neighbourhood and dispersal parameters are 
treated as evolutionary phenotypes: the parameters of the neighbour that 
is doing better are adopted, together with the value of the investment 
amount. Both parameters, as well as the investment amount, are set to 
mutate at a rate of 0.01, following a Gaussian curve with a width 10~\% of 
its peak.

\begin{figure}
\includegraphics{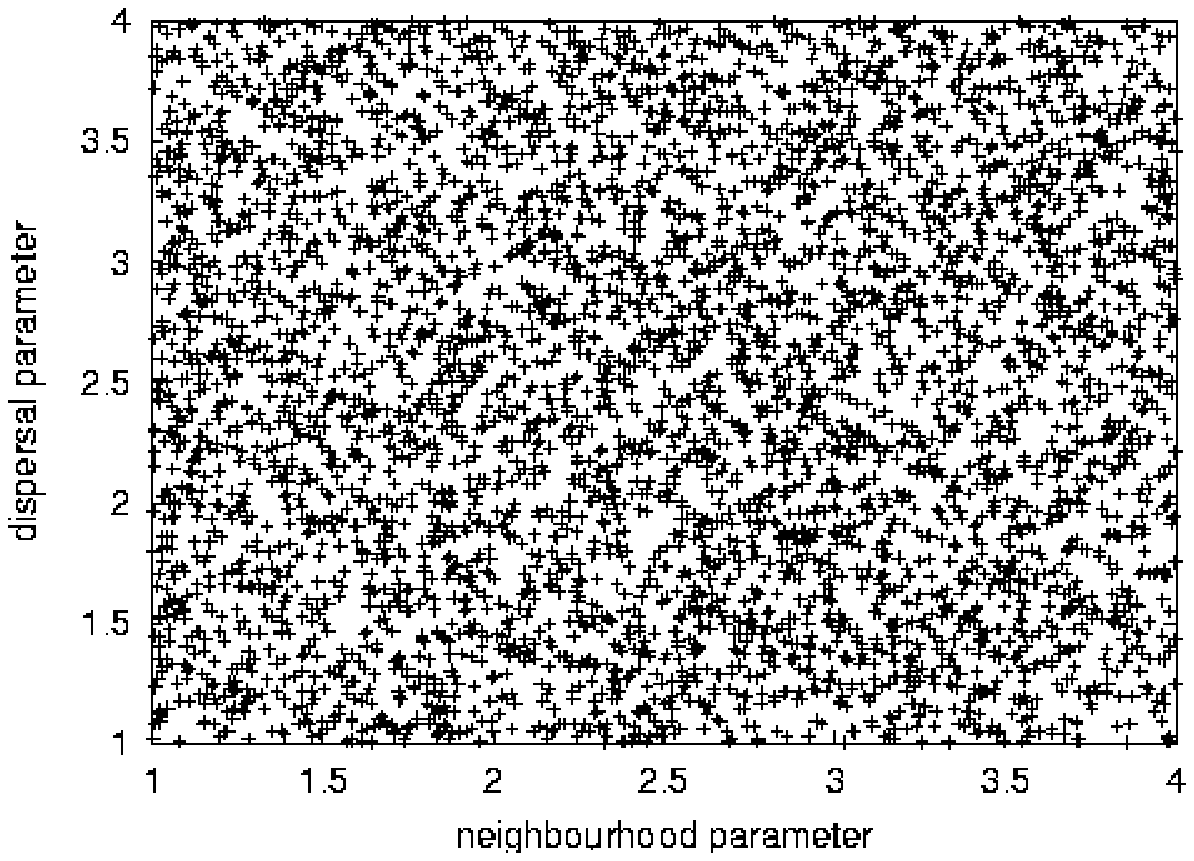}
\includegraphics{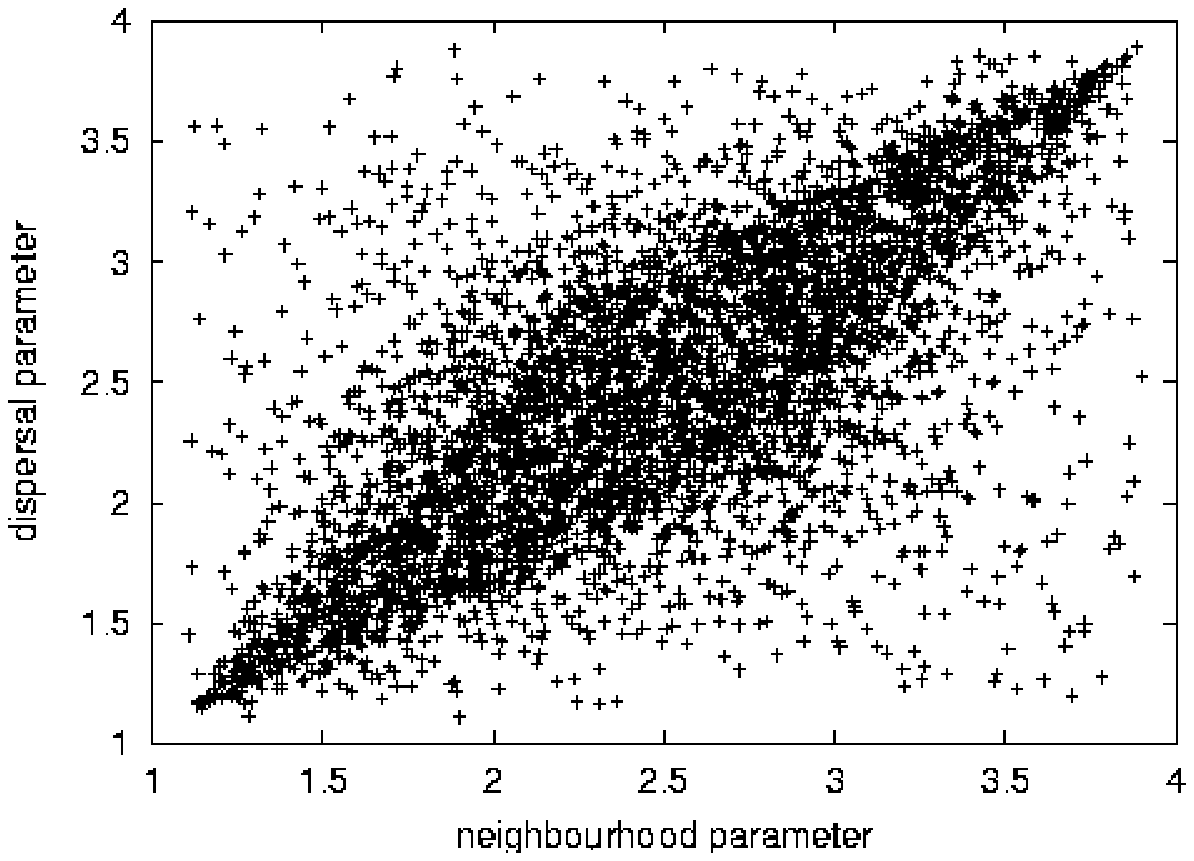}
\caption{(a) At the start of the run the neighbourhood and dispersal
parameters are uniformly distributed between the values 1 and 4. (b) At
the end, they have converged toward one-another. The distribution of 
points is closer to the diagonal $n_i=d_i$.} 
\label{fig:pars}
\end{figure}

\begin{figure}
\includegraphics{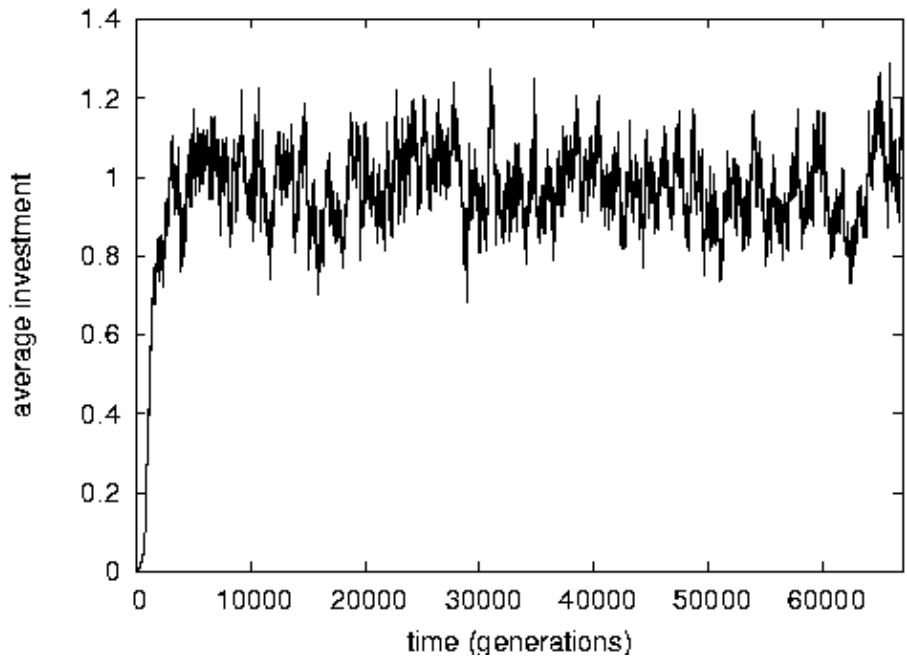}
\caption{The evolution of the average investment when both neighbourhood 
and dispersal parameters mutate, and are passed on from the more
successful individual to the less successful one. Cooperation persists
and investment remains at considerable levels.} 
\label{fig:evol}
\end{figure}

In both scenarios, something interesting happens with the neighbourhood 
and dispersal parameters: in Fig.~\ref{fig:pars} (a) we plot their initial 
values, which are uniformly distributed within a square. At the end of the 
simulation, we find their values have converged closer to the diagonal 
$n_i=d_i$ (see Fig.~\ref{fig:pars} (b)). The convergence of the 
neighbourhood and dispersal parameters toward each-other corresponds to 
the development of a cooperative state, after which the average investment 
continues to sustain relatively high values, almost as high as those 
initially reported by Killingback et al. (1999), but the fluctuations are 
now larger. A typical time series for the average investment in the 
evolutionary scenario is shown in Fig.~\ref{fig:evol}.

\section{CPD in a Network}

Based on the observed dynamics of the spatial CPD on regular lattices, 
Killingback et al. (1999) formulated the hypothesis that the mechanism 
that facilitates and maintains cooperation in worlds of this particular 
topology is their {\it clustering} properties. It is not easy to study 
this hypothesis within the regular lattices topology, since, in order to 
verify or refute it, we would have to obtain results on other topologies, 
different from the regular lattices. Furthermore, for many systems of 
interacting individuals, the regular lattices do not provide a fully 
adequate model for the observed topological properties. These properties 
include large clustering, and short average path length between any two 
nodes (individuals, in our context). 

Watts and Strogatz first introduced ``small world  networks''~(Watts \& 
Strogatz (1998), Watts (1998)), which are clustered structures. A spatial 
structure is called clustered, if the existence of links between nodes A 
and B, and B and C, makes it very probable that there be a link between A 
and C (sociologists call this ``transitivity''). The regular lattices are 
highly clustered structures. They also exhibit a long path length, i.e. 
the number of intermediate nodes between any two nodes A and B is usually 
large. The random graphs of Erd\"os and R\'enyi~(Bollob\'as (1985)), in 
which a node links to any other with a constant probability, have short 
path lengths between nodes, but exhibit no clustering, i.e. are very 
distributed structures. A small world network has the two properties of 
high clustering and short average path length between two nodes, which 
increases with the network size $N$ as the logarithm of $N$~(Watts \& 
Strogatz (1998), Amaral et al.(2000)). Social networks (i.e. networks in 
which humans constitute the nodes, and acquaintances the links) are 
examples of small world networks~(Strogatz (2001), Albert \& Barab\'asi 
(2002), Dorogovtsev \& Mendes (2002)). A special class of small world 
networks are the scale-free ones, in which the degree (i.e. the number of 
the links a node has) distribution obeys a power law (hence it is 
independent of network size). Many algorithms for generating scale-free 
networks have been proposed. This kind of topology is also observed in 
some social networks~(Strogatz (2001), Albert \& Barab\'asi (2002), 
Dorogovtsev \& Mendes (2002), Newman (2001), Barab\'asi et al. (2002), 
Ebel et al. (2002a)). 

Davidsen, Ebel, and Bornholdt~(Davidsen et al.(2002), Ebel et al.(2002b)) 
introduced an algorithm that produces small world networks which follow 
closely the characteristics of social and acquaintance networks. Their 
basic assumption is that the mechanism which generates these networks is 
that people are introduced to one-another by a common acquaintance, 
so-called {\it transitive linking}. At each time step two processes take 
place: $(i)$ A randomly chosen individual introduces two (randomly picked) 
acquaintances of hers to one-another. If this individual has less than two 
acquaintances, she introduces herself to a randomly picked individual. 
$(ii)$ One randomly chosen individual leaves the network with probability 
$p$. All her links are disconnected, and she is replaced by a new 
individual with one randomly picked acquaintance. This probability $p$, 
which is called the death probability, is typically very small. The size
of the network $N$ remains constant in the process. The finite lifetime of
links brings the network to a steady state, in which the degree
distribution is exponential for larger death rates, and becomes power law
(with a cutoff) as the death rate decreases.

The hypothesis that clustering in lattices is the key factor that makes
possible the establishment and maintenance of cooperation can be verified
in networks of different topologies, since some of them exhibit large
clustering, and some do not. To do this, we study the evolution of average 
investment in social networks, and also in random networks, for 
comparison. The two topologies were chosen, since they represent extremes
when clustering properties in networks are considered: the random networks 
are distributed structures with very little clustering, while the social 
networks exhibit a high clustering degree. Also, in these two structures we 
can tune the connectivity (defined as the ratio of the actual number of 
links in the network to the total number of all possible links): for the 
random networks it is equal to the probability of linking between two 
nodes, while for the social networks model it can be varied by varying the 
death probability.

The game proceeds as follows: ``wake-up'' times are generated for each
node as $- \ln (rn)$, where $rn$ is a random number with uniform
distribution in $[0,1]$. The picked node then plays against the nodes it
can ``see'', i.e. it is linked to. Recall that in a network each node can 
have different degrees, i.e. number of immediate neighbours. This means 
that what we previously called neighbourhood parameter differs from node to 
node, and is equal to the degree of the focal node (player). This focal 
player then calculates her payoff from playing against her immediate 
neighbours, according to the rules of the CPD game. These neighbours also 
play against their immediate neighbours, and calculate their respective 
payoffs. The picked (focal) node then compares her playoff to that of the 
neighbour that is doing best, among her immediate neighbours. The 
neighbours she compares herself with (learns from) are the same ones she 
played against (interacted with), i.e. albeit different from one node to 
another, the neighbourhood and dispersal parameters for one node are equal 
to each-other (and the degree of that node). The focal node then adopts the 
investment amount of the neighbour that is doing best as her new investment 
amount, and then the next event is generated. The process is iterated, and 
the average investment is calculated every few ``generations''. This 
process is repeated for different network sizes and averaged over 100 
different configurations for each network size.

\begin{figure}
\includegraphics{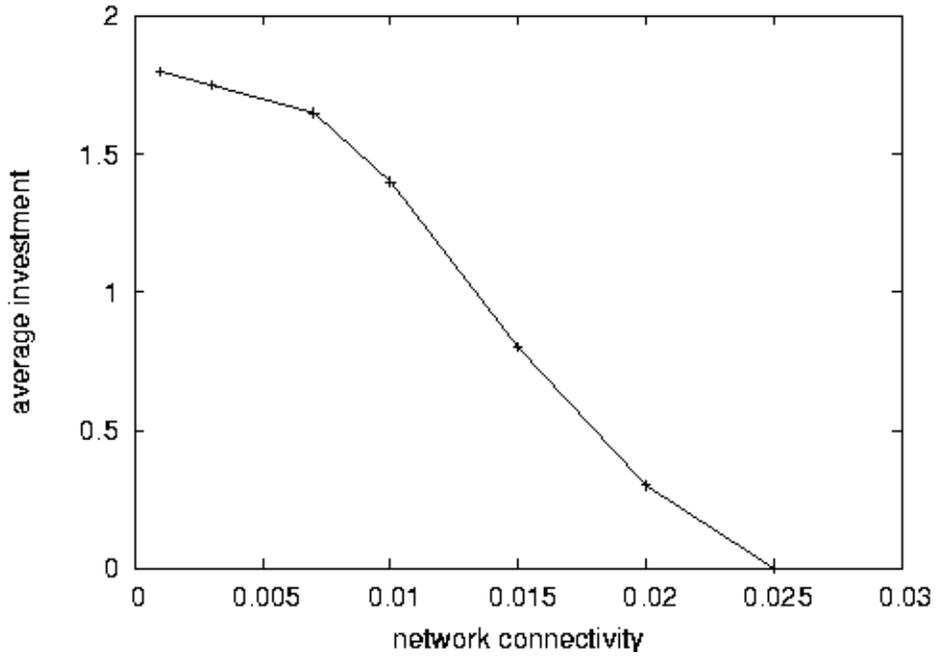}
\caption{The variation of average investment in the steady state with
the connectivity for a social network of size $N=6~000$.}
\label{fig:avinvscc}
\end{figure}

For any fully connected (saturated) network, with connectivity equal to 
one, the final state will be of pure defection, since this is a well-mixed 
system, in which ``everybody sees (or connects to) everybody''. As the 
connectivity decreases, for each network size there is a certain threshold 
value, for which the final state ceases to be that of pure defection, but 
rather a very small average investment. As the connectivity decreases even 
further, the steady state average investment increases. The variation of 
average investment with the connectivity for a social network with 
$N=6~000$ nodes is shown in Fig.~\ref{fig:avinvscc}. The transition from 
the cooperative network to the non-cooperative one is what physicists call 
a critical transition, i.e. the value of the average investment decays 
slowly, as opposed to exhibiting a sharp drop at the crossover connectivity 
value.

\begin{figure}
\includegraphics{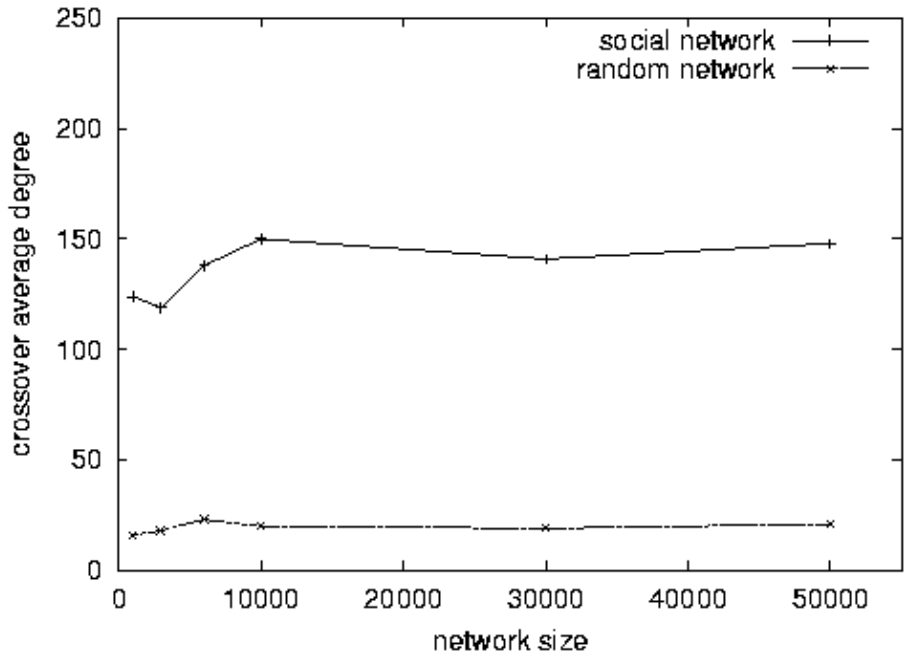}
\caption{The average degree at transition vs. network size. The average
degree remains practically constant, but for the social networks it is
much higher than for the random networks.}
\label{fig:criad}
\end{figure}

For a random network of the same size ($N=6~000$), the same transition is
present, but the connectivity for which the transition happens is 
$C_r=0.003$. This critical connectivity is much smaller than $C_s \simeq 
0.025$ we obtained for a social network of the same size. Since the 
fundamental difference between social and random networks is in their 
clustering properties, it is natural to check the dependence of the 
average degree (number of neighbours) per node at the transition point on 
the network size for both topologies. The results are shown in 
Fig.~\ref{fig:criad}. For both topologies, the critical average degree 
remains essentially constant as the network size is varied. It is about 
150 for the social network model, and about 20 for the random network. 
This surprising result is a very strong argument in favour of the 
hypothesis that clustering is the key mechanism which leads to cooperation 
being first established and then maintained. Indeed, social networks, with 
their pronounced tendency to cluster, maintain the cooperative state for a 
much larger average connectivity (and node degree) than the maximally 
distributed random networks.

\section{Conclusions}

We have studied the behaviour of the Continuous Prisoner's Dilemma model
on different spatial structures, including two-dimensional lattices and 
networks of different topologies. The Continuous Prisoner's Dilemma is an 
evolutionary game which is a natural extension of the standard Prisoner's 
Dilemma. In the Continuous Prisoner's Dilemma each individual makes an 
investment in favour of herself and the neighbours, and the investment 
takes on any value, not just two discrete values. In spatial structures 
like the ones we have considered, and for small neighbourhood 
sizes/connectivities the investment levels evolve by getting higher and 
stabilizing around a given value in a steady state. 

The cooperation develops and is maintained relatively easy, which leads us 
to believe that the cooperation is not such a difficult evolutionary 
paradox. We have tested this in the case of the square lattice, by
increasing the neighbourhood sizes, and found that the neighbourhood size 
for which the steady state is pure defection is about 100 neighbours. As
long as the number of the neighbours the individual plays against
(which we call the neighbourhood parameter) and that of the neighbours the
individual compares herself to (called the dispersal parameter) remains 
the same, the cooperation is quite robust. The mean--field neighbourhood 
size limit does not depend on the size of the lattice, as long as the 
lattice is reasonably large.

Since the interaction neighbourhood and the learning neighbourhood do not 
have to coincide with one-another, we studied the game when the 
neighbourhood and dispersal parameters are different. In this case, we 
found that cooperation only develops for differences between them of about 
0.5 or smaller. Too little information favours defection, but apparently 
so does too much information. Something interesting happens when we allow
the neighbourhood and dispersal parameters to mutate, and at the same 
time, treat them as phenotypes: they are adopted, together with the 
investment amount of the individual that is doing better. The values of 
the neighbourhood and dispersal parameters converge toward one-another, 
i.e. their differences decrease. Cooperation persists, and the average 
investment fluctuates around relatively high levels in the steady state. 

We verified the hypothesis that clustering is the factor that facilitates 
and maintains high average investment values by considering players in 
networks of different topologies. In all of them, a transition is observed 
between the cooperative steady state and the purely defective one, as the 
connectivity of the network increases. For both social networks and random 
ones, the average degree at the transition point is practically 
independent of the network size. However, the average degree at transition 
for social networks is about 150, while that for random networks is much 
lower (about 20). The two topologies differ in their degree of clustering. 
The social networks topology, with its pronounced tendency to cluster, 
allows the establishment and maintenance of a cooperative state, something 
the distributed random networks topology do not. However, when the 
connectivity of the network exceeds a certain value (i.e. too many 
individuals ``see'' one-another and communicate), cooperation can not 
develop or be maintained, and the steady state strategy is pure defection.

The CPD model deepens our understanding as to how cooperation has evolved 
gradually from lower to higher levels in many spatially structured 
systems, where the entities involved can vary from replicating molecules 
to whole organisms~(Wilson (1980), Michod (1983), Buss (1987), Maynard 
Smith \& Szathm\'ary (1995), Dugatkin (1997)): the clustered topology 
helps cooperation develop and survive. It is very encouraging that 
cooperation is quite robust in such systems. On the other hand, in the 
economic world, globalization brings about an increase in the 
connectivities of social and economic networks, and cooperation may 
consequently be less likely to persist.

\section{Acknowledgements}

We thank Birger Bergersen for helpful and interesting discussions and
suggestions.

\newpage

\end{document}